# Statistical mechanics of granular media:
## An approach à la Boltzmann

# P. Evesque

Lab MSSMat, UMR 8579 CNRS, Ecole Centrale Paris
92295 CHATENAY-MALABRY, France, e-mail evesque@mssmat.ecp.fr

### Abstract:

*We use an analogy with the statistical mechanics of gas to build the statistical mechanics of granular media. The case of an isotropic disordered packing of equal spheres submitted to an isotropic stress is considered. We use the assumption of a maximal disorder and the method of the Lagrange multipliers to demonstrate that the distribution of the force moduli decreases exponentially and obeys an exponential-decay statistics. This distribution results from the above hypotheses and from the existence of a mean stress, which acts as a constraint. This distribution is confirmed by experiments and simulations. We apply the same method to find the distribution of void defects of a granular assembly. It follows also an exponential-decay statistics.*

_______________________________________________________________________________

Despite the fact that granular materials are discontinuous materials, their mechanical behaviour is described commonly by continuum-mechanics. This may look ridiculous at first sight, but this works, as proved by the numerous dams and other civil-engineering works which are still in good operating condition and which have been built using this approach.

Besides, this is not the only example when such a continuum approach is applied to discrete media: this is the case of the mechanics of solids and crystals; gas or liquids have been modelled also for long using fluid mechanics. All these media are made of atoms or molecules which are discrete elements. However, what makes the specificity of granular materials compared to the other cases is the absence of Brownian motion and of the analogue of temperature: indeed, it seems that temperature is a good physical ingredient to impose existence of fluctuations which leads in turn to some ergodic principle and which forces the characteristics of the materials to obey some definite macroscopic rules which can be computed from statistical mechanics and which obey some thermodynamics. On the contrary, defining the equivalence of a temperature in granular matter does not seem to be an easy task.

The aim of this paper is to settle the bases of a statistical approach of the mechanics of granular materials. For instance, from an experiment point of view, it has been known from long [1-5] that the amplitude of the forces fluctuates from contact to contact; and this has helped to cast serious doubts on the continuum approach, so that a series of works have been performed already in the 60's and 70's to try and discuss this problem. However, from a theoretical point of view, it is also well known [6] that some macroscopic quantities can be defined from a set of fluctuating quantities. This is the case for instance of the stress tensor $\sigma_{ij}$ [6] which can be calculated from the force distribution $f^c$ and the contact locations $r^c$ in a given volume V.

One of the most important principles which found the theory of statistical mechanics applied to a large number of states is the one which asserts that the observable quantity is that one which corresponds to one of the most probable states. The problem is now to define this most probable state and to calculate its probability and the probability of the others, taking into account the fact that this state is constraint by a series of boundary conditions. This is what we will try to perform in this paper for the mechanics of granular matter.

In a first part, we will briefly recall how this problem has been handled in the case of a gas and will recall the interest of using the Lagrange multiplier method [7] to impose that constraints be satisfied. In the second part, we will apply this method to calculate the statistical properties of the distribution of contact forces in a granular material; for sake of simplicity, this one is assumed to be homogeneous, isotropic and submitted to an isotropic stress; one of the main result is that the force distribution shall decrease exponentially due to the fact that the granular medium is submitted to a well defined stress tensor. The third part will be devoted to calculate the void distribution in a granular medium at a given density.

One of the main interest of this paper is to recall that statistical mechanics is based on the existence of good average quantities which are induced by strong fluctuations at local scale, because the observable states are those ones of the most probable ones, whose probability is much larger than the other states. This is an effect of the large-number law. The system obeys the principle of maximum entropy, or maximum disorder. The second interest is to recall that an exponential decay of the probability of a discrete well-defined quantity results from the existence of constraints at large scale and on the fact that the system is discrete. So, such an exponential decay shall not be viewed as an anomalous behaviour compared to the Gaussian one. It just results from the discrete nature of the system and from the constraints; the Gaussian behaviour being got as soon as the measured quantity is no more a single discrete quantity, but is the sum of a large number of such discrete independent quantities due to the theorem of central-limit.

So, we do not agree with Tkachenko & Witten [8] , when they tell that an exponential distribution of the contact force in granular materials is the demonstration of some anomalous behaviour: it is just the demonstration of the discrete nature of the material; this discrete nature of the system is generated obviously by the existence of contacts, since these contacts condense the stress field into forces at local well defined points and allow the stress to propagate. So, it exists a natural discrete quantity which is imposed by the nature of the materials and its way of working; the paper recalls that the exponential decay is forced simply by the existence of a mean stress.

## 1. Statistical mechanics of gas: the Boltzmann distribution [7]

Let us consider a gas; it is made of N particles which are moving at different speeds $v_\alpha$ with a distribution $p(v_\alpha)$; this corresponds to a kinetic energy $\varepsilon_\alpha = 1/2mv_\alpha{}^2$. Let us consider a large number N of particles and a discrete version of this problem so that the energy $\varepsilon$ is assumed to take discrete values only, i.e. $\varepsilon_i = \varepsilon_o + i\delta\varepsilon$, with i varying from 1 to $\infty$ ($\delta\epsilon$ is assumed to be small compared to the mean value of $\epsilon$, but large enough so that there are a large number $N_i$ of particles which have the energy $\epsilon_i$ in general). So, we are faced to distribute the N particles into the boxes characterised by $\epsilon_i$. Assuming that each particle can be discerned, the complexion number W for a given distribution is:

$$W = N!/ (N_1!…N_i!…) \tag{1a}$$

In statistical physics [7], one assumes that the observed state is the most probable one; this means that it is the state with the largest possible complexion number; so W shall be optimum, and ln(W) too. However, it shall also correspond to a possible state of the system. For instance, it might occur that the state has also a well-defined internal energy U; the total number of particles N is fixed and the total volume V is given. These conditions can be written as:

$$N=\Sigma_i N_i \tag{1b}$$

$$U = \Sigma_i\ \epsilon_i N_i \tag{1c}$$

Or as dU=0, dW=0, dN=0. As from Stirling relation $\ln N! \approx N \ln N - N$, these 3 conditions can be rewritten:

$$d(\ln W)=0= -\Sigma_i \ln(N_i)\ dN_i \tag{2a}$$

$$0= -\Sigma_i\ dN_i \tag{2b}$$

$$0= -\Sigma_i\ \epsilon_i\ dN_i \tag{2c}$$

A way to condense these n=3 conditions into a single one is to use the technique of the Lagrange multipliers and to introduce n-1 coefficients $\alpha$, $\beta$ , …. called the Lagrange multipliers [8]; n is the number of equations to be satisfied at the same time. The conditions (2) become:

$$0= -\Sigma_i \{\ln(N_i)\ -\ \alpha\ -\ \beta\ \epsilon_i\ \}\ dN_i \tag{3}$$

The $dN_i$ are now independent variables and Eq.(3) shall be satisfied without any other constraint, because we have replace Eqs. (2b & 2c) by the knowledge of $\alpha$ and $\beta$. In other word, we have changed of point of view: instead of working with the micro-canonical approach, we are now working with the grand canonical one. Eq. (3) leads to the well known Boltzmann statistical distribution:

$$N_i =\ A \exp\{-\beta\epsilon_i\ \} \tag{4}$$

$\beta$ is 1/kT; it is iposed by U/N since it fixes the width of the distribution of kinetic energy per atom; and A (or $\alpha$) is imposed by the number of particles; both are found by solving Eq. (1b & 1c) after integration.

We propose to use the same method to evaluate the contact-force distribution in a granular medium. So we will assume that the distribution of force obeys some kind of maximum disorder (maximum of entropy) with a specific constraint which warranties that the force distribution respects the boundary conditions and corresponds to a well-given imposed stress field.

## 2. Force distribution of an isotropic granular material :

This part examines the case of a static disordered isotropic homogeneous packing of equal spheres submitted to an isotropic macroscopic stress tensor. It tries and models the distribution of local forces between the grains. From an experimental point of view, it has been known from long [1-5] that the force amplitude fluctuates from contact to contact. From a theoretical point of view, it is also well known [6] that a general relation exists between the force distribution $f^c$, the contact locations $r^c$ and the stress tensor $\sigma_{ij}$ in a given volume V. The stress tensor applied on volume V writes [6]:

$$\sigma_{ij} = (1/V)\sum_{\text{all contacts c in volume V}} f^c_i \, r^c_j \qquad (5)$$

where $f_i$ ,($r_j$), is the component i ,(j), of the contact force ,(contact position c). Then taking into account the facts i) that each grain is in equilibrium ($\Sigma_{\text{force applied on a grain}} f^c = 0$, $\Sigma_{\text{force applied on a grain}} f^c \wedge r^c = 0$), ii) that each grain has the same radius R and diameter D, and iii) taking into account the principle of action and reaction, Eq. (5) can be written as:

$$\sigma_{ij} = (D/V) \sum_{\text{all contacts c in volume V}} f^c_i \, n^c_j \qquad (6)$$

where $n^c_j$ is the component j of the unitary vector $\underline{\mathbf{n}}^c$ defining the direction of the contact c. Considering an isotropic homogeneous sample submitted to an isotropic stress ($\sigma_{ij} = \delta_{ij}$) , and labelling $\underline{\mathbf{u}}^c$ the unit vector in the direction of the force, one can write $\sigma_{ij}$ in the form:

$$\sigma_{ij} = (D/V) \sum_{\text{all possible values of } |f| \text{ in V}} |\mathbf{f}| \{ \sum_{\text{all possible } n^c_i \text{ in V having a given value of } |f|}$$

$$\sum_{\text{all possible } u^c_i \text{ in V, with a defined } |f|} u^c_i \, n^c_j \} \qquad (7)$$

Owing to the symmetry of the system, the first summation is null if $i \neq j$. Let us now assume that the distribution of orientations of $\underline{\mathbf{f}}$ compared to the contact direction is independent of $|\underline{\mathbf{f}}|$ ; in this case, one gets $\langle u^c_i \, n^c_j \rangle = \langle u^c_i \, n^c_i \rangle \delta_{ij} = \delta_{ij} \langle \underline{\mathbf{u}}^c \cdot \underline{\mathbf{n}}^c \rangle / 3 = \delta_{ij} \langle \cos\theta \rangle / 3$ where $\theta$ is the mean angle between the contact direction and the force direction. So, Eq. (7) can be written:

$$\sigma_{ij} = (D/V) \, \delta_{ij} \sum_{\text{all possible values of } |f| \text{ in the volume V}} |\mathbf{f}| \, \langle \cos\theta \rangle / 3 \qquad (8)$$

Eq. (8) can be rewritten:

$$\sigma_{ij} = D \langle \cos\theta \rangle / (3V) \, \delta_{ij} \sum_{\text{all possible values of } |f| \text{ in the volume V}} |\mathbf{f}| \, N_{|\mathbf{f}|} \qquad (9a)$$

where $N_{|\mathbf{f}|}$ is the number of contacts whose force modulus is in between $|\mathbf{f}|$ and $|\mathbf{f}| + |d\mathbf{f}|$. We proceed now as in the case of a gas and assume a maximum of disorder (principle

of maximum entropy); so, the complexion number shall be maximum, but the total number of contacts be constant. This leads to the two other Equations:

$$\ln(W) = \ln(N!) - \sum_{\text{all possible values of } |\mathbf{f}| \text{ in the volume V}} \ln(N_{|\mathbf{f}|}!) \qquad (9b)$$

$$N = \sum_{\text{all possible values of } |\mathbf{f}| \text{ in the volume V}} N_{|\mathbf{f}|} \qquad (9c)$$

After derivation of Eqs.(9), one can solve this system of equation using the method of Lagrange multiplier; one gets:

$$N_{|\mathbf{f}|} = A \exp\{-\beta|\mathbf{f}|\} \qquad (10)$$

Reporting Eq. (10) in Eqs (9a & 9c) allows to calculate A and b as a function of $\sigma_{11}+\sigma_{22}+\sigma_{33}=Tr(\mathbf{s})$ and of the mean number of contacts per unit volume n=N/V; this leads to the probability distribution:

$$p(|\mathbf{f}|) = N_{|\mathbf{f}|}/N = [nD\langle\cos\theta\rangle/Tr(\mathbf{s})] \exp\{-n|\mathbf{f}|D\langle\cos\theta\rangle/Tr(\mathbf{s})\} \qquad (10)$$

Indeed this probability distribution and its exponential tail seem to be in agreement with recent numerical results [8-10], with experimental data [11-12]. It leads to similar results as the scalar model of [13] in a much simpler way.

It is worth noting that results (10) will remain the same if one have assumed that all contacts where at sliding; in this case $\langle\cos\theta\rangle$ would be simply $\cos\varphi$, with $\varphi$ being the angle of solid fiction.

Of course, this approach may be oversimplified; for instance, it does not take into account the process of building; this one might influence the dependence of $\langle\cos\theta\rangle$ as a function of the force modulus $|\mathbf{f}|$ and might change the probability distribution. Also existence of local correlation between forces might affect the whole distribution. However we shall remark that Eq. (6) does not depends explicitly on the correlations on between forces on different contacts, but just on local force and local distance of the contact to the centre of gravity of each grain; so, if the correlation length of the force is finite, we shall expect that force-force correlation be unimportant on the global law, since it shall only affect the size of the discrete quantity to be used. So, it seems that it is only in the case when the correlation length would be infinite that this approach should be affected importantly; but in that case we should observe some kind of order and the material behaviour should affected at a macroscopic scale also.

Anyhow, it seems that the exponential tail is a result of an optimisation which tends to broaden the distribution while respecting the value of the mean stress. It is possible that L. Rothenburg has argued in a similar way in his PhD. Dissertation [11].

## 3. void distribution in a granular assembly:

In the same way, we may evaluate the void distribution in a granular assembly. Let us define the void index $e=V_{voids}/V_{solid}$ as the volume of voids $V_{voids}$ divided by the volume of solids $V_{solid}$, and the porosity $\phi = V_{voids}/V_{tot}$ as the void volume divided by the total volume $V_{tot}$. As $V_{tot}=V_{voids}+V_{solids}$, one gets the relation $\phi = e/(1+e)$. Let us also consider the grains as rigid. In this case any granular material contains voids; and it exist a maximum packing density, which will be characterised by $e_{min}$; oviously $e_{min}$ depends on the distribution of the size and the shape of the grains. In the same way, in absence of cohesion, this granular material exhibits a minimum density characterised by $e_{max}$, and $e_{max}$ depends on the size and the shape of grains. At an intermediate density $e_{min} < e < e_{max}$ the material exhibits some disorder linked to the introduction of additionnal voids. Let us consider that this new voids are linked to the introduction of some kind of discrete defects, each of these defects being characterised by a given void volume v and let us try to evaluate the distribution $N_v$ of the number of voids of a given volume v. We start in analysing the problem in the microcanonical approach. This means that the total number of defects $N = \Sigma \, N_v$ and void volume $(e-e_{min})V_{sol} = \Sigma \, vN_v$ are fixed.

So a reasoning similar to the ones in sections 1 and 2, which assumes a maximum of the entropy, leads to evaluate the probability of finding a defect of volume v to follow a law of the kind:

$$p(v) = N_v/N = A\exp\{-\beta v) \tag{11}$$

Assuming now that the typical size of a defect is the size of a grain $v_o$, one gets from the two conditions $N = \Sigma \, N_v$ and of void volume $(e-e_{min})V_{sol} = \Sigma \, vN_v$ :

$$p(v) = [(e-e_{min})v_o)]^{-1} \ \exp\{-v/[(e-e_{min})v_o)]\} \tag{12}$$

This means that the distribution of size of the defect follows an exponential-decay law, as it was assumed by Boutreux and de Gennes [14] in a paper describing the compaction due to cyclic hitting. A future paper will try and adapt their approach to the problem of the natural densification of a granular assembly due to increase of mean pressure; the variation of void index as a function of the pressure increase will be determined from this microscopic modelling.

## 4. Conclusion

One of the main interest of this paper is to recall that statistical mechanics is based on the existence of good average quantities which are induced by strong fluctuations at local scale, because the observable states are those ones of the most probable ones, whose probability is much larger than the other states. This is an effect of the large-number law. The system obeys the principle of maximum entropy, or maximum disorder. It seems that such an approach was tempted already by L. Rothenburg in his PhD. Dissertation [11]. The second interest is to recall that an exponential decay of the probability of a discrete well-defined quantity results from the existence of constraints at large scale and on the fact that the system is discrete. So, such an exponential decay shall not be viewed as an anomalous behaviour compared to the Gaussian one. It just results from the discrete nature of the system and from the constraints; the Gaussian behaviour being got as soon as the measured quantity is no more a single discrete quantity, but is the sum of a large number of such discrete independent quantities due to the theorem of central-limit.

So, we do not agree with Tkachenko & Witten [8] , when they tell that an exponential distribution of the contact force in granular materials is the demonstration of some anomalous behaviour: it is just the demonstration of the discrete nature of the material; this discrete nature of the system is generated obviously by the existence of contacts, since these contacts condense the stress field into forces at local well defined points and allow the stress to propagate. So, it exists a natural discrete quantity which is imposed by the nature of the materials and its way of working; the paper recalls that the exponential decay is forced simply by the existence of a mean stress.

Assuming that an increase of void index is linked to the generation of void defects, we were able to find the statistics of these defects using the principle of maximum of entropy. This has allowed to find the exponential distribution which has been used by Boutreux and de Gennes in their modelling of densification by cyclic hitting.

*Acknoledgments:* CNES is thanked for partial funding.

The electronic arXiv.org version of this paper has been settled during a stay at the Kavli Institute of Theoretical Physics of the University of California at Santa Barbara (KITP-UCSB), in june 2005, supported in part by the National Science Fundation under Grant n° PHY99-07949.

*Poudres & Grains* can be found at :
http://www.mssmat.ecp.fr/rubrique.php3?id_rubrique=402